\documentstyle[aps,rotating,prc,graphicx]{revtex}
\begin{document}
  \title {One-body dissipation and chaotic dynamics \\
    in a classical simulation of a nuclear gas}

  \author{M. Baldo, G. F. Burgio, A. Rapisarda}

  \address{Istituto Nazionale di Fisica Nucleare, Sezione di
    Catania and Dipartimento di Fisica, Universit\'a di Catania, Corso
    Italia 57, I-95129 Catania, Italy } 

  \author{P. Schuck}

\address{Institut de Physique Nucl\'eaire, Universit\'e de Grenoble, 
53 Avenue des Martyrs, 38026 Grenoble Cedex, France}

\maketitle

\bigskip

\begin{abstract} 
  In order to understand the origin of one-body dissipation in nuclei,
  we analyze the behavior of a gas of classical particles moving in a
  two-dimensional cavity with nuclear dimensions.  This "nuclear"
  billiard has multipole-deformed walls which undergo periodic shape
  oscillations.  We demonstrate that a single particle Hamiltonian
  containing coupling terms between the particles' motion and the
  collective coordinate induces a chaotic dynamics for any
  multipolarity, independently on the geometry of the billiard. If the
  coupling terms are switched off the "wall formula" predictions are
  recovered.  We discuss the dissipative behavior of the wall motion and
  its relation with the order-to-chaos transition in the dynamics of the
  microscopic degrees of freedom.
\end{abstract}

\bigskip
\bigskip

{PACS number(s): 24.60.Lz, 21.10.Re}

\bigskip

In the last twenty years the dissipation of collective motion in nuclei
has been widely observed \cite{sp81} in low energy particle and heavy
ion collisions and it still represents a theoretically unsolved problem.
It is commonly believed that both one-body processes, {\it i.e.}
collisions of nucleons with the nuclear wall generated by the common
self-consistent mean field, and two-body collisions produce dissipation,
although their interplay is not well known \cite{sme,bort}.\par
The theory of damping and the approach to equilibrium of a collective
(usually slow) degree of freedom coupled to non-collective (usually
fast) degrees of freedom is a quite general one, both at classical and
quantal levels. In the last decades several papers have been
published on the subject, see for example refs. 
\cite{wilk,Jar,Ber,Bul,wilki,blum,bswi,wf,weid,pal,bauer,bbrs}.
The most studied case is the extreme adiabatic limit,
namely when the ratio between the characteristic times of the fast
degrees of freedom and of the slow one is considered arbitrary small.
Under the further assumption that the fast degrees of freedom have an
ergodic dynamics for a fixed value of the collective variable, a
description of the damping and approach to quasi-equilibrium can be
given in terms of generalized diffusion equations \cite{wilk,Jar}.
The theoretical framework for the treatment of this limiting case,
at least at the classical level,
seems to be well established, despite some questions appear to be still 
open \cite{Ber}.
Unfortunately, in the majority of realistic cases the adiabatic
condition is only partly satisfied and the ergodic condition is not
often fulfilled. The damping, however, is still necessarily connected to some
degree of chaoticity of the overall dynamics of the system. The
connection between chaotic dynamics and dissipation in this more general
situation has been studied by different authors but
still lacks of a general theoretical framework.
In particular Wilkinson \cite{wilk} presented some evidences 
which indicate that the occurrence of an integrable or nearly
integrable motion of the fast degrees of freedom strongly
suppresses both the fluctuations and the speed of relaxation
towards equilibrium. Related studies have been developed in ref. \cite{latora},
where a coexistence of chaoticity and slow 
relaxation to equilibrium in a Hamiltonian Mean Field (HMF) model has
been found, and in ref. \cite{dellago}, where the 
relation between chaoticity and the approach to equilibrium in a hard
sphere gas and a Lorentz gas has been analyzed.

In this context, Blocki {\it et al.}\cite{bswi} analyzed the behavior
of a gas of classical non-interacting particles enclosed in a
multipole-deformed container which undergoes periodic shape
oscillations. Particles move on linear trajectories and collide
elastically against the walls. In this model wall and particles' motion
is uncoupled and therefore the wall keeps oscillating at the same
frequency pumping energy into the gas.  For this system, the authors
study the increase of the particles' kinetic energy as a function of
time.  They find that for octupole and higher modes the gas kinetic
energy increases with time, in agreement with the "wall formula"
predictions \cite{wf}.  They attribute the different behavior to the
fact that for low multipolarity deformation the particles' motion is
regular and corresponds to an integrable situation, whereas for higher
multipolarities the shape irregularities lead to divergence between
trajectories and therefore to chaotic motion.  Although their results
look very interesting, their application to the nuclear or similar cases
is not straightforward because i) the self-consistent mean field is
absent, ii) the total energy is not conserved.\par
A step forward in this direction has been performed by Bauer {\it et
al.} in ref.\cite{bauer}. In this work the authors study the damping
of collective motion in nuclei within the semiclassical Vlasov equation.
Here selfconsistency is taken into account and the total energy is
conserved. A multipole-multipole interaction of the Bohr-Mottelson type
is adopted for quadrupole and octupole deformation.  In both cases the
dynamical evolution shows a regular undamped collective motion which
coexists with a weakly chaotic single-particle dynamics.\par
In ref.\cite{bbrs} we introduced a simple model which can shed light on
the correlation between chaoticity and dissipation.  We considered a gas
of classical non-interacting particles moving in a two-dimensional
billiard with nuclear dimensions. Particles collide with the oscillating
walls and transfer energy to it, but the container can give back this
energy heating the gas.  We considered the gas + billiard as an
Hamiltonian system, therefore the total energy is conserved with a good
accuracy. Though no explicit dissipative term is considered the main
effect of the coupling of the wall with the particles is a damping of
the collective motion.  The particle - wall coupling considered in our 
model is relevant for at
least two other reasons : i) the coupling can enhance collectivity of
the motion, since the particles are indirectly coupled among each others
through the wall, ii) only with coupling included the motion of the
particles can be driven to equilibrium at large timescales.  In fact, it
has been shown \cite{JarSw} that without the coupling the asymptotic
particle velocity distribution is non-maxwellian.\par
In ref.\cite{bbrs} only the monopole case was
considered together with a small number of particles, namely 1 and 10.
Here we extend the previous study by considering more particles (N=30)
and L=2,3 multipolarities. 
For a fixed value of the wall deformation, the motion of the particles
is regular in the case of the monopole L=0, partly chaotic for
large deformation in the quadrupole case L=2, and essentially
chaotic in the octupole case L=3. In this way we are exploring
a set of physical conditions for which the motion of the fast degrees 
of freedom ranges from nearly ergodic to integrable (for a
fixed value of the slow variable). The parameters of the model are
chosen as typical at the nuclear scales. However, the considered
set of dynamical systems should be generic enough to be 
representative of the physical problem under study, namely the
damping process of a slow degrees of freedom in a bath of many
fast degrees of freedom, far enough from adiabaticity and for
different degrees of ergodicity. We therefore expect that the
results we have obtained are generic enough to be qualitatively
valid for a wide class of physical systems.\par
The most important findings of our study are: a) chaos shows up in the single
particle motion for any surface deformation for long time scales; b) the
different geometry influences only the timescale for the onset of
chaoticity which is faster for the higher multipolarities; c) the
different timescales for the onset of chaoticity  are in any case equal 
within a
factor of two; d) the dissipation of the collective motion is sensitive
only to this timescale and thus it depends only
slightly on the geometry of the billiard for L=2 and L=3; e) though no 
explicit dissipative term is considered in our model, the damping of the
collective motion is in practice irreversible due to the large Poincare'
time.\par
The present paper is organized as follows. We summarize the details of the
model in section 1. The numerical results are illustrated in section 2.
Conclusions are drawn in section 3.

\section {The Model}
In ref.\cite{bbrs} we considered a classical version of the vibrating
potential model (see {\it e.g.} ref.\cite{risc}). In this model several
non-interacting classical particles move in a two-dimensional deep
potential well and hit the oscillating surface. Using polar coordinates,
the Hamiltonian depends on a set of $\{r_i, \theta_i \}$ variables,
describing the motion of the particles, and the collective coordinate
$\alpha$.  The Hamiltonian reads

\begin{equation}
H(r_i, \theta_i, \alpha) = \sum_{i=1}^N ({{p_{r_i}^2}\over {2m}} +  
{{p_{\theta_i}^2} \over{2mr_i^2}} +
V(r_i, R(\theta_i))) + {{p_\alpha^2} \over{2M}} + {1\over 2}
M \Omega^2 \alpha^2   \label{eq:model}
\end{equation}
\noindent
where $\{p_{r_i}, p_{\theta_i}, p_{\alpha}\}$ are the conjugate momenta
of $\{r_i, \theta_i, \alpha\}$.  $m = 938~MeV$ is the nucleon mass, and
$M = \eta m N R_o^2$ is the Inglis mass, chosen proportional to the
total number of particles $N$ and to the square of the circular billiard
radius $R_o$.  The value of the factor $\eta$ is fixed in such a way to
minimize the equilibrium fluctuations.  In our case $\eta = 1$ for the
monopole oscillation, whereas $\eta = 10$ for quadrupole and octupole.
Therefore in the $L=0$ case, collisions of particles against the walls
are more inelastic.
\noindent
$\Omega$ is the oscillation frequency of the collective variable
$\alpha$. The potential $V(r, R(\theta))$ is zero inside the billiard
and a very steeply rising function on the surface, ${V(r, R(\theta))} =
{V_o \over {(1 + exp({{ R(\theta) - r } \over {a}}))}}$, with $V_o =
1500~MeV$ and $a = 0.01~fm$.  Such a small value of the diffuseness
was chosen in order to simulate closely a billiard system. Larger
values of $a$ should not affect qualitatively the results.
The surface is described by $R(\theta) =
R_o (1 + \alpha~ cos(L \theta))$.  Therefore this potential couples the
collective variable motion to the particles' dynamics and prevents
particles from escaping.  The numerical simulation is based on the
Hamilton's equations

\begin{equation}
\dot r_i = {p_{r_i} \over m},~~~ \dot p_{r_i} = {p_{\theta_i}^2 
\over {m r_i^3}} - {\partial V \over \partial r_i}
\end{equation}

\begin{equation}
\dot \theta_i = {p_{\theta_i} \over {m r_i^2}},~~~ \dot p_{\theta_i} 
= - {\partial V \over \partial R} {\partial R \over \partial \theta_i}
\end{equation}

\begin{equation}
\dot \alpha = {p_{\alpha} \over M},~~~ \dot p_{\alpha} = 
-M \Omega^2 \alpha - \sum_i ({\partial V \over \partial R_i}
{\partial R_i \over \partial \alpha})
\end{equation}
\noindent
Please note that dropping the V term in the equation for $p_\alpha$, the 
wall motion becomes purely harmonic and decoupled by the particles' motion,
as assumed in the model proposed by the authors of ref. \cite{bswi,wf}.

We solve the Hamilton's equations with an algorithm of fourth-order
Runge-Kutta type with typical time step sizes of $1~fm/c$.  If not
otherwise stated, the calculations were performed with the number of
particles $N = 30$. The total energy was conserved with relative error
$\Delta E/E \leq 10^{-4}$.  For runs with the longest time duration or
with the largest number of particles a fourth order symplectic
integrator \cite{Yo} was used, since it turns out to be more efficient
with comparable degree of accuracy.\par
As far as the initial conditions are concerned, we assign random
positions to the particles inside the billiard and random initial
momenta according to a two-dimensional Maxwell-Boltzmann distribution
with a temperature $T = 36~MeV$. This value of temperature is chosen
in order to mimic the Fermi motion of the particles. In this way the 
average velocity of the particles is close to the typical nuclear
Fermi velocity. In a classical description the use of a Maxwellian 
distribution ensures that the initial conditions are not too far
from equilibrium for the whole system.

We consider the wall oscillations taking place not too far from adiabaticity.  
For this purpose we follow the definition of the adiabaticity parameter
$\kappa$ given by the authors of ref.\cite{bswi}, {\it i.e.}

\begin{equation}
\kappa = \frac {\alpha_o \Omega R_o} {v}
\end{equation}

being $v$ the most probable particle speed, $v = \sqrt {T/m}$, $\alpha_o$
the initial amplitude of the oscillation and $R_o$ the radius of the
circular billiard. We choose $R_o = 6~fm$ and $\Omega = 0.05~s^{-1}$
so that 

\begin{equation}
\kappa = 1.53 \alpha_o 
\end{equation}

With our choice of $\alpha_o$ (see below), the adiabaticity parameter
$\kappa$ might be not as
small as assumed in the asymptotic theory of ref. \cite{wilk,Jar}. The
adopted value can correspond more to realistic situations.\par
Since in the realistic cases the collective motion takes place around
equilibrium, the initial wall coordinate has been chosen equal to
$\alpha_0 = \bar\alpha + \delta \alpha$, where $\bar \alpha$ is the
equilibrium value and $\delta \alpha$ the deviation.  The equilibrium
value $\bar \alpha$ corresponds of course to the thermodynamic limit,
which is actually reached when considering an ensemble of copies of the
system, all of them with an initial value $\alpha = \bar \alpha$ and
differing from each other in the initial microscopic distribution of
particles' positions and momenta. The equilibrium value depends on the
considered multipolarity.  In all cases we checked numerically that
starting from the thermodynamical value $\alpha = \bar \alpha$ and a
Maxwell-Boltzmann distribution, the collective variable $\alpha$
oscillates in time around $\bar \alpha$. Furthermore, we considered an
ensemble of this type of runs, each one corresponding to different
particles' initial conditions, consistent with a space uniform
distribution and a Maxwell-Boltzmann velocity distribution. We then
checked that the average over the runs of the value of collective
variable $\alpha(t)$, at any given time $t$, was converging indeed to
the constant equilibrium value $\bar \alpha$ for a reasonable large
number of runs. In this way one can also estimate the fluctuations
around equilibrium the variable $\alpha$ undergoes in a typical run.
Also the average amplitude of the fluctuations turns out to be
consistent with the thermodynamical estimate for a harmonic oscillator
in a thermal bath. More details on the procedure can be found in ref.
\cite{bbrs}.  After we checked that the numerical simulation produces
good equilibrium properties, we perturbed the equilibrium collective
coordinate $\bar \alpha$ by an amount $\delta \alpha = 0.3$, and let
both the billiard and the particles evolve in time.  The chosen value is
larger by about a factor of 3 than the average equilibrium fluctuations.
Moreover at time $t=0$ we put $p_\alpha = 0$, the wall having only
potential energy.

\section {Numerical results}
\subsection {Scatter plots}
One possible way in order to investigate the role played by the coupling
and see whether it can induce chaotic dynamics, is drawing Poincare's
surface of sections for the single particle coordinate.  However this is
impossible to perform in our case because of the large number of degrees
of freedom. Then an alternative way to visualize a chaotic behavior is
to draw scatter plots, see Fig.1.  There we display the final radial
coordinate at a time $t$ of one particle vs. the one at $t=0$ for
the monopole, quadrupole and octupole deformation. The chosen times are
$t=0.4 \tau$, $1.6 \tau$, $2.4 \tau$ and $4\tau$, $\tau$ being the period 
of the oscillation, $\tau = \frac{2\pi}{\Omega}$.  
These plots are very useful and are
commonly used in transient irregular situations like chaotic scattering
\cite{scatt} and nuclear multifragmentation \cite{frag}. The idea is
that if the dynamics is regular, two initially close points in space
stay close even at later times, but if the dynamics is chaotic the two
points will soon separate due to the exponential divergence induced by
chaos.  In the first case this plot will show a regular curve, whereas
in the other one a diffused pattern appears.  We note that for all
multipolarities the initially regular curves change into scattered dots,
which clearly show that the coupling to the wall oscillation randomizes
the single particle motion. This is at variance with what was discussed
in ref.\cite{bswi}, where chaos is supposed to appear only for
multipolarities $L > 2$.  In our model the coupling between wall and
particles' motion produces a chaotic dynamics even for $L \leq 2$.  In
addition, the higher the multipolarity the earlier chaos starts because
of the increased shape irregularity.

\begin{figure}
 \begin{center}
\includegraphics[bb= 0 0 515 719,angle=90,scale=0.5]{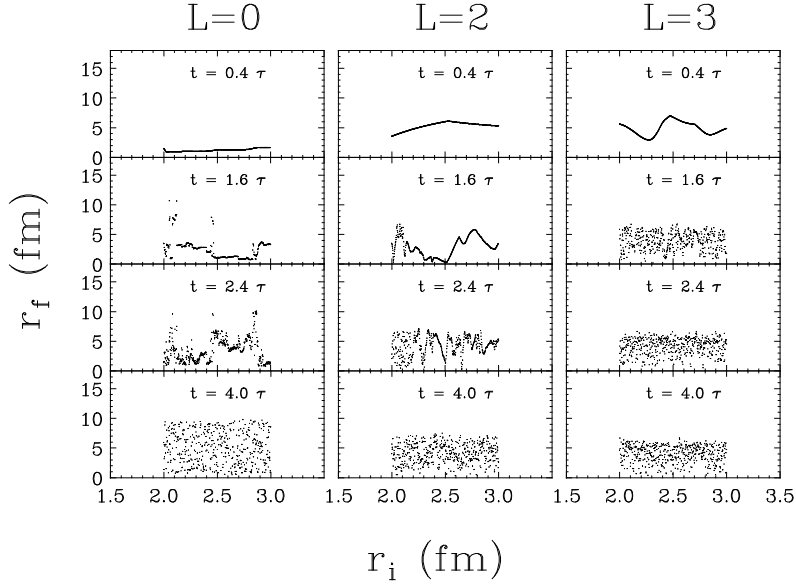}
\end{center}
\caption[]{\footnotesize {The final radial coordinate for one particle is
  drawn as a function of the initial one at different times t = 
$0.4 \tau$, $1.6 \tau$, $2.4 \tau$ and $4 \tau$. Calculations are 
performed for multipolarities L = 0,
  2, 3. 1000 initial conditions are considered.}} \label{fig:scat_plot}
\end{figure}

\begin{figure}
\begin{center}
\includegraphics[bb= 0 0 515 719,angle=90,scale=0.5]{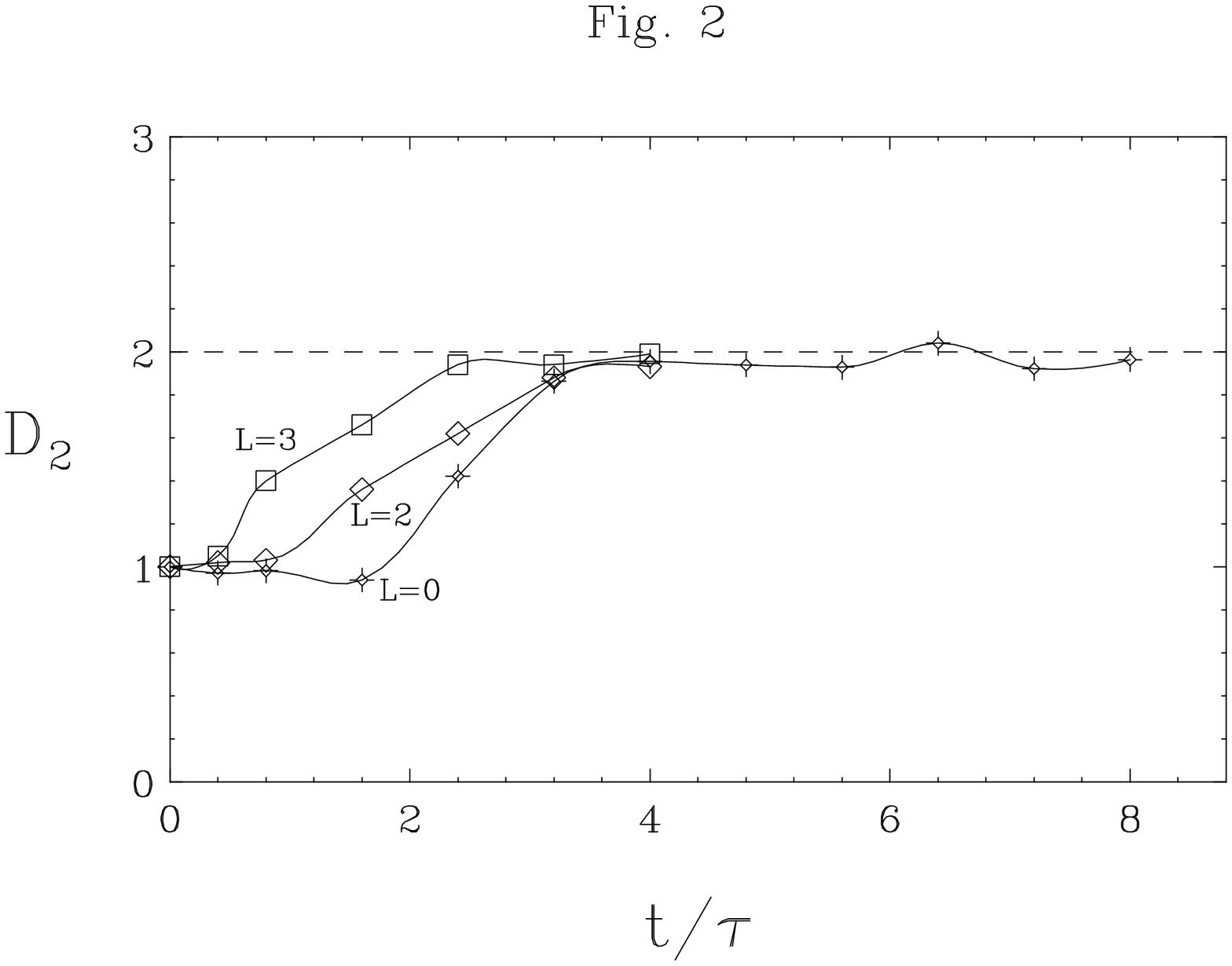}
\end{center}
\caption[]{\footnotesize {The time evolution of the fractal correlation 
  dimension $D_2$ is plotted for the multipolarities L = 0, 2, 3.  The
  lines are to guide the eye. The dashed line is the fully random limit
  for scattered points on a plane.}} \label{fig:corr}
\end{figure}

A more quantitative analysis can be performed because scatter plots of
Fig.1 have a typical fractal structure. As already done in
ref.\cite{frag}, a fractal correlation dimension $D_2$ can be calculated
from the correlation integral $C(r)$ \cite{grass}. One first counts how
many points have a smaller distance than some given distance $r$. As $r$
varies, so does the correlation integral $C(r)$, defined as

\begin{equation}
C(r)~ = ~ {1\over {M^2}} \sum_{i,j}^M \Theta(r - |{\bf z_i} - {\bf z_j}|),
\label{eq:corr}
\end{equation}
where $\Theta$ is the Heaviside step function and ${\bf z_i}$ a vector
whose two components ($x_i, y_i$) are the points coordinates.  M is the
total number of points.  The fractal correlation dimension $D_2$ is then
defined by

\begin{equation}
D_2~ = ~\lim_{r\to~0}~ {ln~C(r) \over {ln~r}}.
\label{eq:frac}
\end{equation}

Therefore by plotting the logarithms it is possible to extract $D_2$ by
fitting the linear slope for sufficiently small r.  We considered as a
good interval satisfying eq.(\ref{eq:frac}) the one between $r_{min} =
1.83\cdot 10^{-2} $ and $r_{max} = 1 $.  An ensemble of 1000 points was
considered.
In Fig.2 we display $D_2$ vs. time for each multipolarity.  At the very
beginning $D_2$ is equal to one, showing that the motion is regular.  As
time goes on, regularity is lost and the motion becomes chaotic until a
complete randomness is reached, in which case $D_2~=~2$ as expected for
a completely random distribution on the plane\cite{frag}.  This result
confirms the one published in ref.\cite{bbrs}, where a different method
of calculation was however employed.

\subsection{Lyapunov exponents}
A more common quantitative way to characterize the dynamics in a chaotic
regime is by means of the largest Lyapunov exponent $\lambda_1$, which
gives the average rate of exponential divergence of two trajectories
with nearly identical initial conditions. In general, if we denote with
$d(t)$ the distance between two phase space trajectories and $d_o =
d(0)$ , $\lambda_1$ characterizes the average growth of the distance $d(t)$ 
with time

\begin{equation}
<d(t)>~ =~ d_o ~ exp(\lambda_1 t).
\end{equation}

\begin{figure}
\begin{center}
\includegraphics[bb= 0 0 515 719,angle=0,scale=0.5]{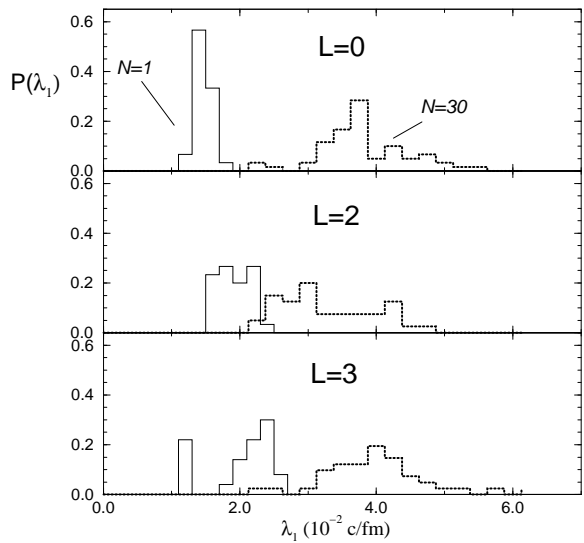}
\end{center}
\caption[]{\footnotesize {For the cases L=0,2,3 we show the frequency 
distributions 
  of the largest Lyapunov exponent $\lambda_1$ for 1 (thin histogram)
  and 30 particles (thick histogram). An ensemble of about 40 events was
  considered in each case.}}\label{fig:lyap}
\end{figure}

We have calculated this largest Lyapunov exponent $\lambda_1$ by the
method of Benettin {\it et al.} \cite{ben}. This method consists in
evolving two close trajectories originally separated in phase space by
$d_o$, for a given time interval $\tau$, after which the magnitude of
their separation $d(\tau)$ is rescaled back to $d_o$. The procedure is
then repeated k times. It can be shown that $\lambda_1$ is given by the
following limiting procedure :

\begin{equation}
\lambda_1 =  \lim_{k \rightarrow \infty} ~~ \lim_{d_o \rightarrow 0} 
{1 \over {k \tau}}~~ \sum_{i = 1}^{k} ln{d_i(\tau) \over d_o}
\end{equation}

The result turns out to be essentially independent of $d_o$.  We
considered the following metric in phase space

\begin{equation}
d(t) = \sqrt{\sum_{j=1}^N (\delta x_j^2 + \delta p_{x_j}^2 
+ \delta \theta_j^2 + \delta p_{\theta_j}^2 )}  
\end{equation}

where the infinitesimal distances where normalized to the mean values
and therefore are dimensionless quantities.  In our case the length of
the time interval $\tau$ was 200 fm/c and $d_o \simeq 2. 10^{-6}$.
Using the above method, a large number of trajectories (of the order of
40) was sampled for each multipolarity. Each trajectory was followed for
a time $t = 3500 fm/c$, thus allowing for a safe determination of the
largest Lyapunov exponent.

In Fig.3 we display for each multipolarity a distribution of $\lambda_1$
for a number of trajectories and for 1 and 30 particles (respectively
thin and thick histogram). The most probable values do not depend
strongly on the multipolarity, at variance with the results found in
ref.\cite{bswi} and lie within a factor of 2 one from each other, the
case $L=0$ being intermediate between $L=2$ and $L=3$.  This is probably
due to the different choice of the wall mass.  In the case N=1,
$\lambda_1$ is always smaller than the value obtained with N=30, thus
showing that a) chaos starts earlier when a higher number of particles
is present and b) for long timescales the degree of chaoticity is more
dependent on the number of particles than on the multipolarity.
Therefore an Hamiltonian with coupling terms gives rise to a dynamics
weakly dependent on the geometry of the billiard.  It has to be noticed
that the average number of particle-wall collisions per oscillation
period is only about 2 for N=1 and about 60 for N=30.

\subsection {Chaos and dissipation}

\subsubsection{Analysis of the single event}

In this section we will mainly discuss the interplay between chaoticity
of the single particle dynamics and dissipation of the collective
motion. For this purpose, we follow at the same time the dynamics of the
wall and the particles.

\begin{figure}
\begin{center}
\includegraphics[bb= 0 0 515 719,angle=90,scale=0.5]{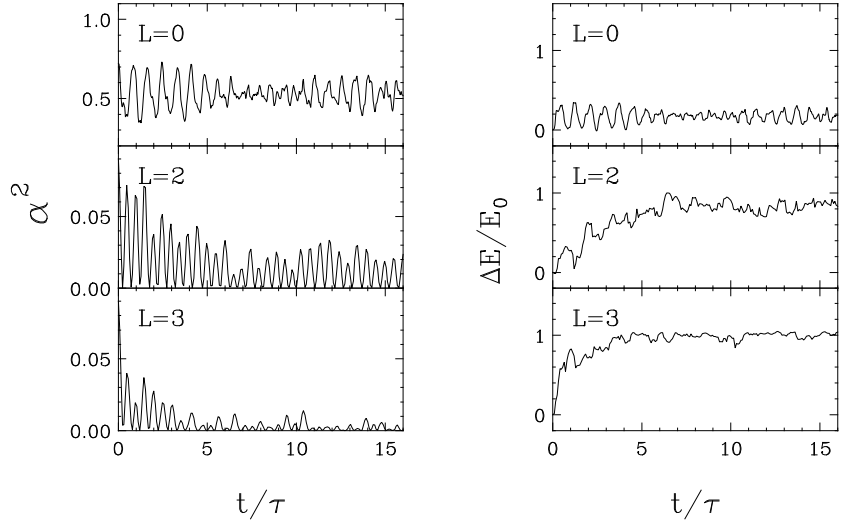}
\end{center}
\caption[]{\footnotesize {On the left-hand side the time evolution of the
  collective variable is shown for the multipolarities L = 0, 2, 3 and
  on the right-hand side the corresponding excitation energy of the gas,
  normalized to the initial energy $E_o$.}}\label{fig:1_event}
\end{figure}

First we analyze the behaviour of one single event, keeping in mind that
a correct statistical description can be performed only for an ensemble
of events, as already pointed out in \cite{bbrs}.  On the left-hand side
of Fig.4 we plot the evolution of the square of the collective variable
$\alpha^2$ vs. time for one single event. Each panel corresponds to a
fixed multipolarity. We note that the amplitude of the collective motion
shows an irregular oscillation, at variance with the results found in
ref.\cite{bauer}.  We note also that a slight damping can be observed,
in contrast with the results published in ref.\cite{bbrs}, where
calculations with a smaller number of particles were performed.  Please
note that $\alpha^2$ keeps on oscillating around the equilibrium value
$\bar \alpha$ which for the monopole case depends on the gas temperature
and on the wall frequency \cite{bbrs}. For the other multipolarities
$\alpha$ oscillates around zero and $\alpha^2$ decreases more rapidly
for increasing L, indicating a faster dissipation for a larger
irregularity of the billiard.

\begin{figure}
\begin{center}
\includegraphics[bb= 0 0 515 719,angle=90,scale=0.5]{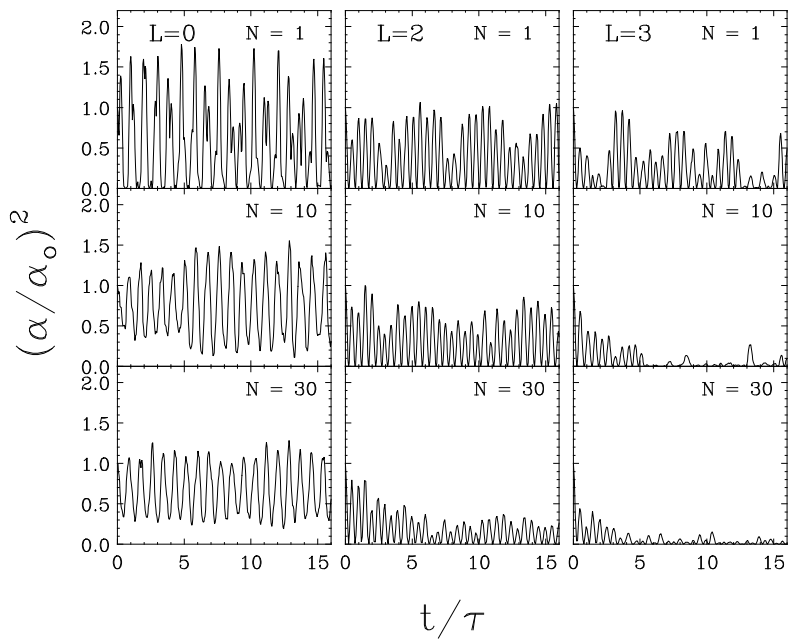}
\end{center}
\caption[]{\footnotesize{The square of the ratio $\alpha/\alpha_o$ is plotted
  vs. the number of oscillations for different multipolarities and 
number of particles.}}
\label{fig:1_event_a}
\end{figure}

On the right-hand side of Fig.4 we plot the time evolution of the
excitation energy of the gas, defined as the relative variation of the
total energy of the gas $E$ with respect to its initial value $E_0$,
$\Delta E/E_0$. In all three cases the gas is heated up, but with a
different time dependence according to the multipolarity. Except for
small irregular fluctuations, an increasing trend shows up for the
quadrupole and octupole modes. An oscillating pattern, slightly
irregular, is visible for the monopole case.  Moreover, in the $L=0$
oscillation the energy gained by the gas is lower than in the $L=2, 3$
cases. Therefore some dissipation is present for all multipolarities and
is larger for increasing $L$. Of course these are only general features
of the events for different multipolarities, the details being different
from one event to the other.\par
In order to better understand the mechanism of dissipation, we display
in Fig.5 the square of the ratio $\alpha / \alpha_o$, being $\alpha_o =
\alpha(t=0)$. The plot is as function of time, for different
multipolarities L and particle number N. In our model the total energy
is conserved so the damping of the wall motion corresponds to the
heating of the gas.  This process is in principle reversible, but the
high number of degrees of freedom involved makes in practice this
process irreversible.  One should wait a very long time, $i.e.$ the
Poincare' time, to see the system again close to the initial conditions.
We note that the dissipation is stronger when both L and N increase. In
particular, the increased number of particles makes the available phase
space bigger and the Poincare' time longer.  The latter can therefore be
a limitation, in the sense that a short Poincare' time would weaken
dissipation.

\subsubsection{Analysis of an ensemble of events}
 
In order to have a global picture of the macroscopic system, many events
are needed and average values of the different observables should be
considered.  In Fig.6 we display the time evolution of $\alpha^2$
averaged over an ensemble of 1000 different events, each obtained by
assigning random initial conditions to the particles both in coordinate
and momentum space but consistent with a Maxwellian.  We note that the
motion of the collective variable $< \alpha^2 >$ is completely damped out
for times which depend on the multipolarity and develops around
equilibrium, as we checked explicitly.  The reader should keep in mind
that for $L=0$ collisions of the particles with the wall are more
inelastic (see paragraph 2) due to the lighter mass of the wall,
therefore the monopole oscillation damps out in a time comparable with
the one of the $L=2$ mode.  The same mass was used in the calculations
published in ref.\cite{bbrs}.  If we put $\eta = 10$ even for $L=0$, the
damping time is longer.

\begin{figure}
\begin{center}
\includegraphics[bb= 0 0 515 719,angle=90,scale=0.5]{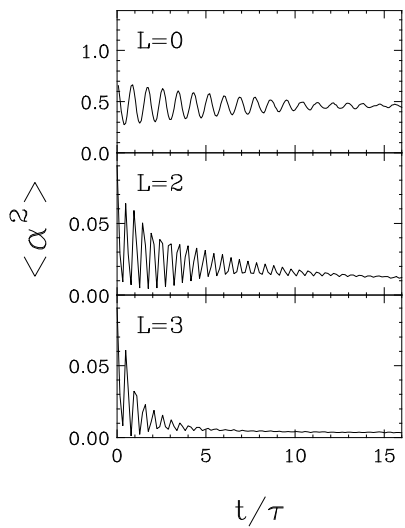}
\end{center}
\caption[]{\footnotesize {The square of the collective variable $\alpha$, 
averaged  over an ensemble of 1000 events, is displayed vs. the number 
of oscillations for the three
  multipolarities considered.}}\label{fig:ens_event}
\end{figure}

\begin{figure}
\begin{center}
\includegraphics[bb= 0 0 515 719,angle=90,scale=0.5]{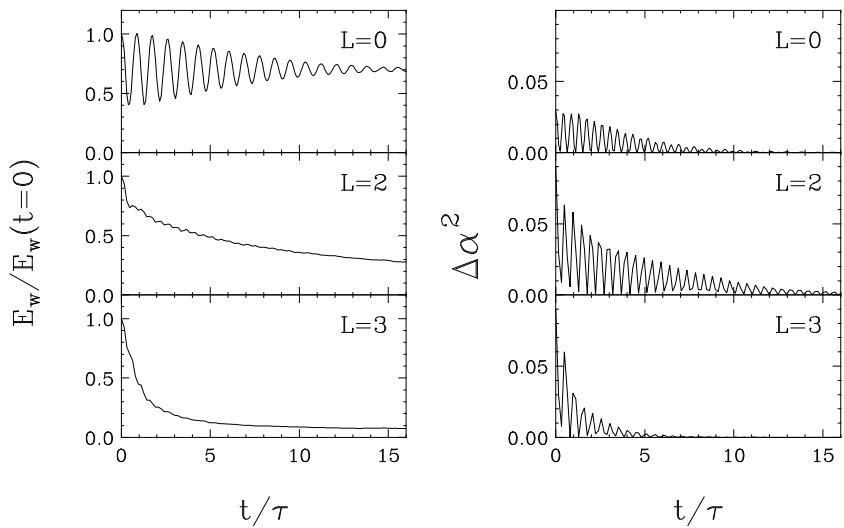}
\end{center}
\caption[]{\footnotesize {On the left-hand side the wall energy, averaged 
over an ensemble of 
  1000 events and normalized by its initial value, is reported vs. the 
  number of oscillations
  for different multipolarities. On the right-hand side the observable
  $\Delta\alpha^2$ is drawn (see text for further details).}}
\label{fig:wall}
\end{figure}

The damping of the oscillations, which is apparent in Fig. 6 for all the
three multipolarities, can have two distinct origins.  On one hand, each
individual event can indeed display a damping of the oscillation
amplitude. In this case the effect of the average is only of smoothing
the fluctuations. The observed damping time is then simply the average
damping time of a generic set of events.  On the other hand, each event
has a different time evolution, and after a period of time the {\it
  phase} of the oscillation can be quite different from one event to
another. If the phase becomes essentially random as the time proceeds,
when taking the average of $\alpha$ over different events strong
cancellations can occur and a damping of the oscillations can appear,
even in the case where no average energy damping is present. It is worth
noticing that this damping mechanism disappears if the ratio between the
single particle mass and the wall mass is vanishing small, and therefore
it is ineffective at macroscopic level, where event-to-event
fluctuations can be neglected.  In order to estimate the contribution of
each one of the two mechanisms to the damping displayed in Fig. 6, we
plot on the left-hand side of Fig. 7 the energy of the wall divided by
its initial value, averaged over the same sets of events.  As a function
of time, this quantity should be insensitive to phase randomization
because it is proportional to $<\alpha^2>$ and corresponds to the
average energy damping of the oscillations.

For comparison, on the right-hand side we plot also $\Delta\alpha^2 \ =\ 
(<\alpha> -\alpha_\infty)^2$, a quantity which, on the contrary, should
be sensitive to phase cancellations.  For the monopole case, the decay
time of the wall energy to its asymptotic value $\alpha_\infty$
is substantially longer
than the one for $\Delta\alpha^2$, which indicates that for the monopole
case the dephasing mechanism is quite important.  The asymptotic value
of $<\alpha>$ is consistent with the equilibrium value $\bar\alpha$
for a gas at the
final temperature, which, from energy conservation, turns out to be $T =
43~MeV$. We checked that indeed the velocity distribution, averaged over
the events, is very close to a Maxwellian with that temperature.
Furthermore the average amplitude of the fluctuations around the
equilibrium value was checked to be consistent with a harmonic
oscillator in equilibrium with a thermal bath at $T =~ 43~ MeV$.\par
On the contrary, for the higher multipolarities the decay time of the
wall energy and $\Delta\alpha^2$ appear to be very close, and therefore
the dephasing mechanism seems to be in these cases ineffective. The
damping is mainly an energy damping and is characteristic of a generic
event.

\begin{figure}
\begin{center}
\includegraphics[bb= 0 0 515 719,angle=90,scale=0.5]{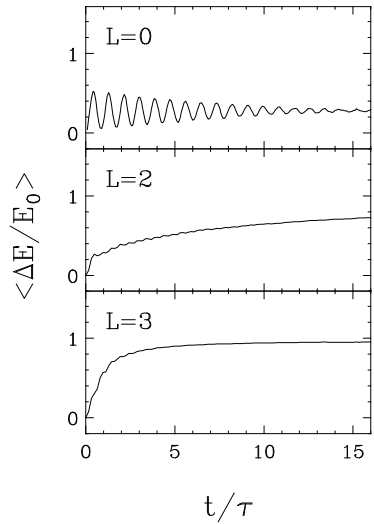}
\end{center}
\caption[]{\footnotesize {The excitation energy of the gas, averaged over 
an ensemble of 1000 events, is reported vs. the number of oscillations 
for different multipolarities.}}
\label{fig:gas}
\end{figure}

These considerations are confirmed if one calculates the excitation
energy of the gas, displayed in Fig.8. We note that the characteristic
times for the energy growing follow closely in all cases the decay times
of the wall energy.  Moreover for $L=2, 3$ we note the presence of two
different regimes: a first one lasting for about one two oscillations 
characterized by a sharp rising in the excitation energy, and a second
one at successive times where a slower approach to equilibrium is
apparent.  As it can be deduced from Fig.1, the first stage may have
some relation with the onset of chaos in the single-particle motion.
After chaos has fully developed, and an appreciable part of the total
energy has been pumped into the gas, a certain degree of randomization
is reached and a slower dissipation rate shows up. A similar behavior
in the relaxation was observed for the HMF model \cite{latora} and
the hard sphere gas and the Lorentz gas studied in ref.\cite{dellago}.
In the latter cases the time evolution of the Boltzmann entropy was studied.

This behavior is completely absent in the monopole case. 
Therefore it seems that
different kinds of dissipation can originate from the same underlying
chaotic single-particle motion.  It should be stressed that, within the
time of chaos development, the wall has dissipated only a fraction of
its energy, as it is clearly seen in Fig. 7.  In the monopole case it
is apparent that the flow of energy from the collective variable
$\alpha$ to the particle gas is not completely ``irreversible'', and the
wall receives back part of its energy from the gas for a number of
oscillations.

\subsubsection{Comparison with the wall formula}

Let us now try to make some connection to one body dissipation and the wall 
formula introduced and studied extensively in the past by Swiatecki and 
coworkers \cite{bswi,wf}. We limit ourselves to the analysis of the 
octupole mode, because for that (and for higher multipolarities)
the Swiatecki model predicts dissipation. 
Of course a comparison between the situation 
studied here and in \cite{bswi} may be biased because in the model studied
in \cite{bswi} there is no energy conservation. 
Therefore phase space of the gas is 
conserved in the latter model but not in the case considered here. 
In spite of that our model should be able to recover the Swiatecki limit. 
And actually it does. This happens when in our model the mass of the wall 
becomes very large as for example in Fig.9 where we choose $\eta = 100$ and 
$\alpha_o = 0.09$ with $N = 30$ and $\Omega = 0.05$ as before. 
In this case the particle-wall collisions are more elastic and the time 
evolution is slower for all observables. In particular in panels (c,d)
of Fig.9 the dots are our numerical simulation and the solid and dashed lines
are two versions of the wall formula to be explained in the 
appendix. In particular, the dashed line represents the results of the 
wall formula which contains only terms linear in time. The inclusion 
of terms quadratic in time, which take into account for the increased
average speed of the gas for large times, produces a new dissipation 
formula represented in the figures below by a solid line. 
Now, if every thing is consistent we should get the same result as
the one shown in Fig.9 in switching off the coupling. That this is 
actually the case is shown in Fig.10 where the time evolution of the 
different quantities is practically the same as in Fig.9.

\begin{figure}
\begin{center}
\includegraphics[bb= 0 0 515 719,angle=90,scale=0.5]{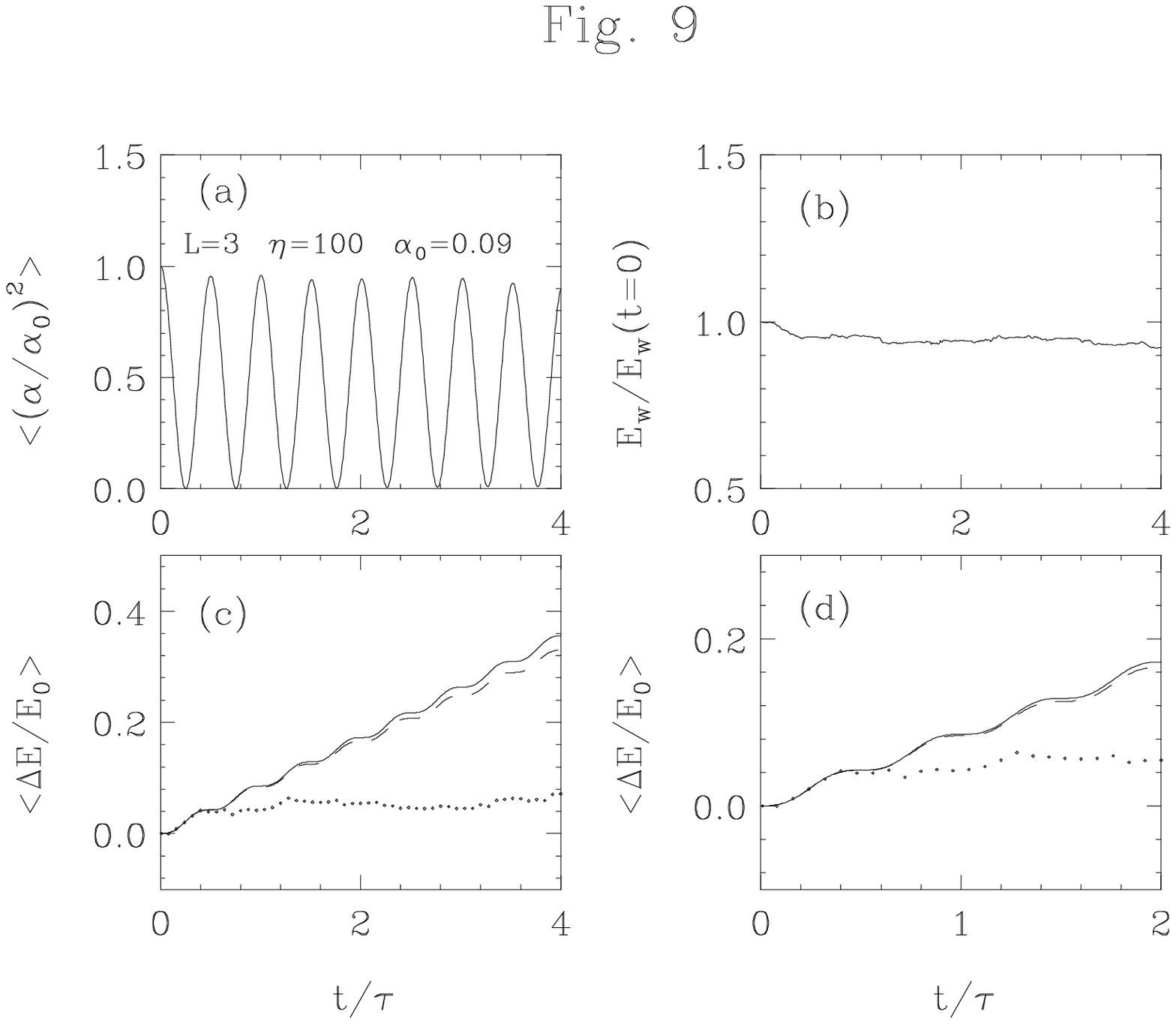}
\end{center}
\caption[]{\footnotesize {Coupled octupole motion of the wall and gas particles
with a very large inertia of the wall ($\eta = 100$). The adiabaticity
parameter is $\kappa = 0.14$. Shown are the elongation (panel (a)), 
the wall energy (panel (b)), the relative increase of the gas energy 
(panel(c)), and a blow up of panel (c) (panel(d)).}}
\label{fig:schuck9}
\end{figure}

\begin{figure}
\begin{center}
\includegraphics[bb= 0 0 515 719,angle=90,scale=0.5]{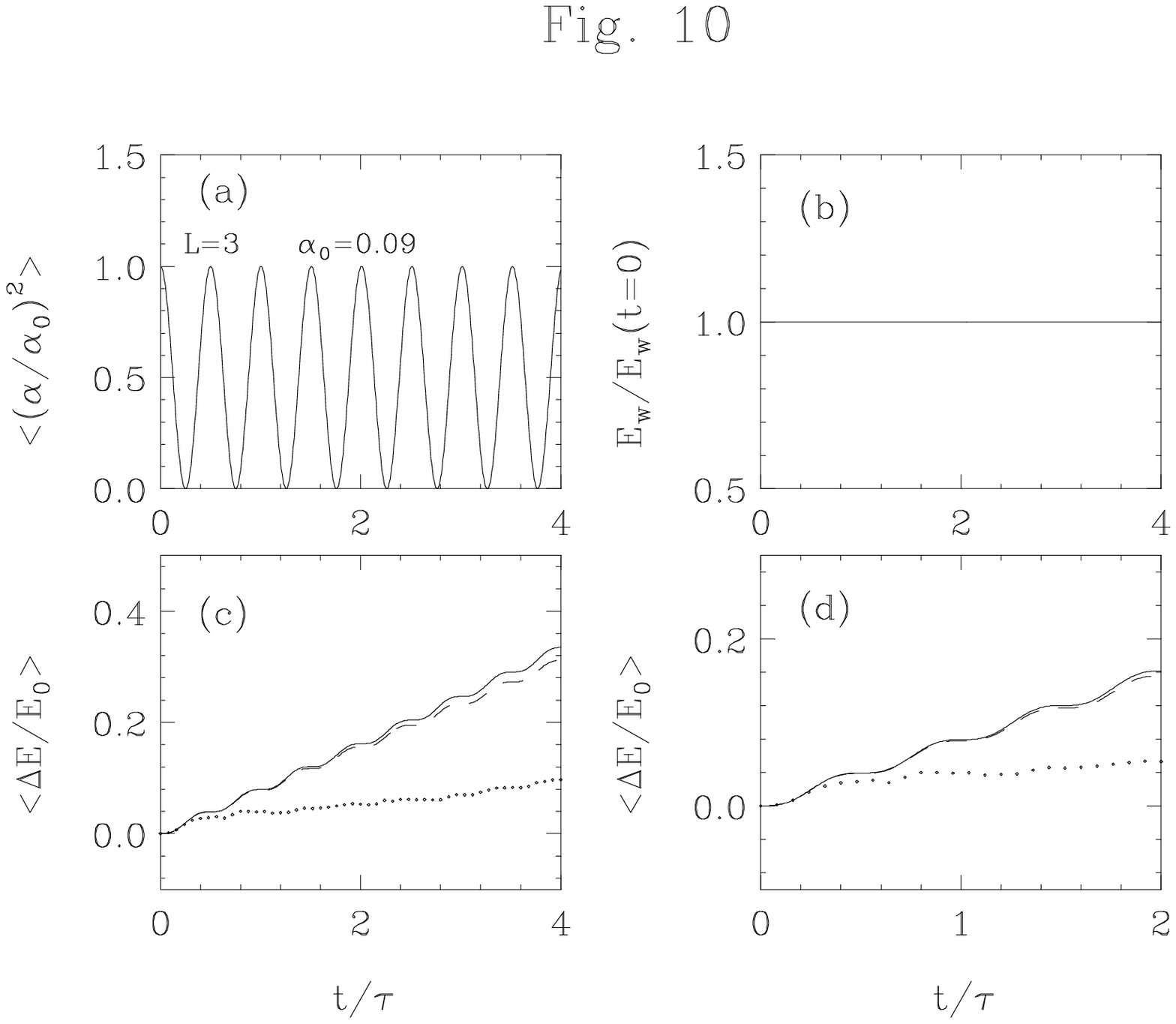}
\end{center}
\caption[]{\footnotesize {Same as Fig. 9 but without coupling. The 
frequency of the wall is kept fixed to the initial one, {\it i.e.}
$\Omega = 0.05 s^{-1}$.}}
\label{fig:schuck10}
\end{figure}

A particular feature of the case shown in Figs. 9-10 is the fact that the wall
formula results follow the exact evolution only during the first period. 
We should emphasize that we applied here the wall formula adapted to the 
two dimensional case, as derived in the appendix. In Fig. 11 we show a 
repetition of the calculation in ref.\cite{bswi} where the wall formula
follows the exact result for the heating of the particles gas over 
several periods. The only difference with the calculation in \cite{bswi}
is that we here use again for the gas particles our Maxwell-Boltzmann 
distribution with $T=36~ MeV$, whereas in \cite{bswi} a sharp Fermi-Dirac 
distribution was employed. The fact that the wall formula for this 
particular case of the parameters is well reproduced indicates that this 
feature is independent of Maxwell-Boltzmann or Fermi-Dirac statistics.
This must be expected, since in the wall formula the dissipation rate depends
only on the average velocity of the particles. Therefore, the dissipation
must be the same as long
as the average velocity of the particles is the same. The only variance with 
the case shown in Figs. 9-10 is that the parameters are slightly different,
or that for instance the adiabaticity parameter $\kappa$ defined in 
eq.(6) is by almost a factor of 4 smaller in Fig.11 than in Figs. 9-10, 
that is
$\kappa = 0.04$ instead of $\kappa = 0.14$. However, both values of
$\kappa$ are sufficiently small in order to verify the adiabaticity criterion
(for the validity of the wall formula) established by Swiatecki et al., 
namely that $\kappa << 1$. According to our findings here, $\kappa$ must 
really be extremely small so that the wall formula holds during several
periods, otherwise it may be valid only during the first instances of the
dynamics.

\begin{figure}
\begin{center}
\includegraphics[bb= 0 0 515 719,angle=90,scale=0.5]{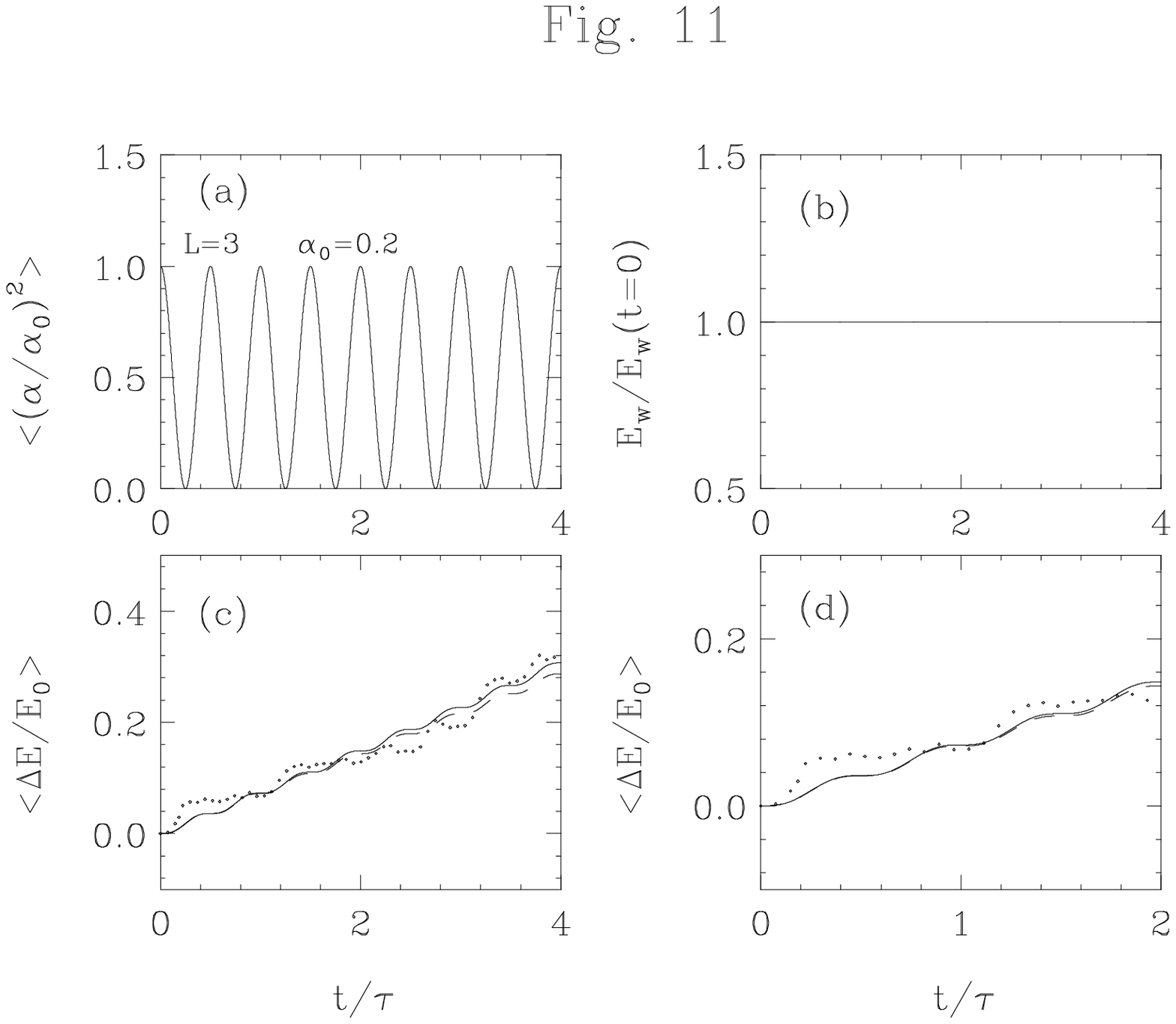}
\end{center}
\caption[]{\footnotesize {Repetition of Swiatecki et al. calculation but with
a Maxwellian distribution for the gas particles. 
The observables displayed are the same as in Figs.9-10.
The adiabaticity parameter is $\kappa = 0.04$. See the text for more details.}}
\label{fig:schuck11}
\end{figure}

Let us now come back to a more detailed analysis of our present results
obtained including the coupling terms in the Hamiltonian.
In Fig.12 we show as a function of time the evolution of the elongation 
(panel (a)), the wall energy (panel (b)), the gain in energy of the gas
(panel (c)) together with a blowup of (c) (panel (d)). In panels (c) and (d)
are also shown the results of the standard wall formula (broken line), 
which is linear in time \cite{bswi}, 
and the new version including a quadratic term (full line) \cite{wf2}.
We should stress that we applied here generalized wall formulas taking 
account of the fact that the wall motion is actually damped. This implies 
that we have to use a time dependent version of the wall formula (TDWF)
as derived in the appendix.

\begin{figure}
\begin{center}
\includegraphics[bb= 0 0 515 719,angle=90,scale=0.5]{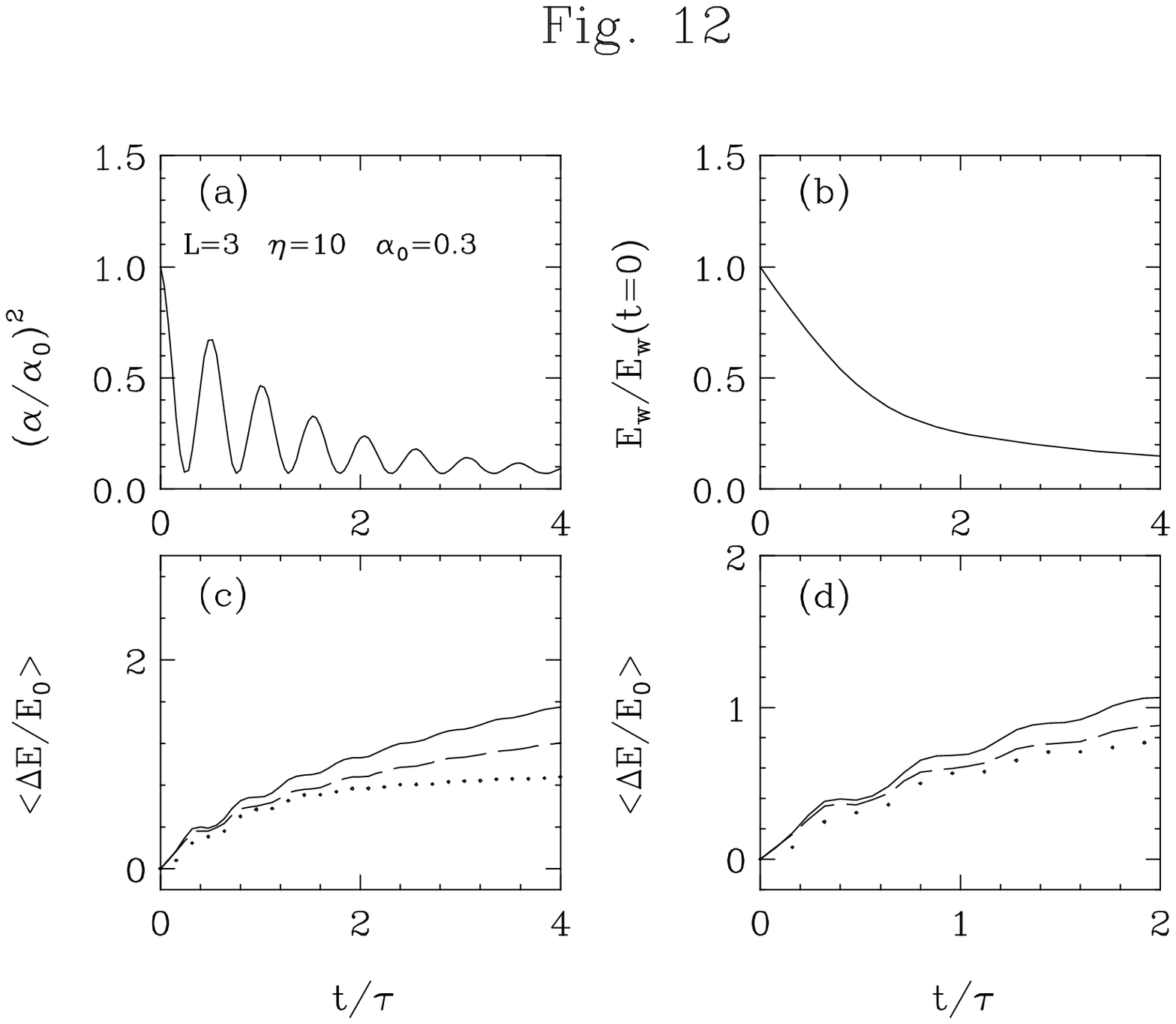}
\end{center}
\caption[]{\footnotesize {Same as Fig.9 with coupling. 
In panels (c) and (d) time dependent wall formulas have been used, 
taking into account respectively terms linear in time (broken line) with
quadratic ones (full line). The dots are our numerical results.}}
\label{fig:schuck12}
\end{figure}

Again we see that the wall formula only agrees with the exact solution during 
the first period. For longer times the wall formulas give rise to much
more energy dissipation than in the exact evolution, the new formula more
than the standard one. 

From the above investigations it seems therefore that the wall formula can 
only simulate the dissipation of energy in the first instances of the 
dynamics when  the adiabaticity parameter $\kappa$ (eq.(6)) is
extremely small (of the order of 1/100).

\section {Conclusions}

In conclusions, we have presented a dynamical approach based on the solution
of the Hamilton's equations for several particles moving in a classical
billiard having nuclear-like dimensions, in order to explain dissipation
of the collective motion. We found that the presence of a coupling term
in the single particle Hamiltonian induces chaotic motion at the
microscopic level. \par
As far as the monopole mode is concerned, we found irregular behavior
together with damping in single events.  The damping observed when an
average is taken over a set of events is a consequence also of the
irregular time dependence of the oscillations.  This incoherence is
produced by the chaotic single particle dynamics, which makes all events
belonging to the same ensemble different one from each other, and
therefore substantial cancellations occur once the average is taken.
For the higher multipolarities this mechanism seems to be ineffective,
and the damping coincides with a real energy damping of the
oscillations due to the large Poincare' time.  
The dissipative process for the quadrupole and octupole
modes looks different also in another respect.  In fact, while the
single event properties are similar to the monopole case, an ensemble of
events shows that two different regimes appear : a) an initial fast
dissipative evolution corresponding to the onset of chaos in the
single-particle motion and b) a slower dissipative trend towards
equilibrium. \par
All these results should be qualitatively independent on the particular
nuclear dimensions we have used in the model.\par
In order to be closer to equilibrium we have used for the initialisation 
of the gas particles a Maxwell-Boltzmann distribution with a temperature
of $T=36~MeV$ to have a mean kinetic energy characteristic of nuclear
systems. We also checked that the use of a Maxwell-Boltzmann distribution
versus a Fermi-Dirac step is apparently of no consequence on the damping
rate. Indeed in switching off the particle wall coupling and periodically
octupole-deforming the wall under the same condition as was done some time
ago by Swiatecki and collaborators \cite{bswi} using a Fermi-Dirac step, we
find with a Maxwell-Boltzmann distribution an identical rate of feeding of
energy into the particle gas. Moreover we also checked that the wall
formula prediction for energy dissipation agrees with the numerical
simulation over several periods in time. This shows that our model is 
able to mock up real nuclear dynamics together with its damping mechanisms.

When we increase the adiabaticity parameter $\kappa$ by almost 
a factor of four 
attaining $\kappa = 0.14$, what should still be considerable as small, 
the agreement of the numerical result with the wall formula was 
reduced to only the first period of the oscillation. In view of what we said 
above about Maxwell-Boltzmann versus Fermi-Dirac distributions, we assume 
that this is a generic result. Also when reducing the inertia of the wall
by a factor of ten and thus increasing the particle wall coupling when
reestablished the heating rate of the particle gas agrees with the wall 
formula only during the first period. 
This happens in spite of the fact that we used a time
dependent version of the wall formula where the continuous slowing down
of the wall motion has been taken into account. Indeed the wall formula 
considerably overestimates the damping rate at longer times.

One should realise that our model implies real particle-wall collisions with 
global energy conservation. Such collisions are absent in pure Hartree-Fock
or Vlasov calculations. In fact our model in what concerns the particle
wall collisions should come quite close to the situation considered
{\it e.g.} in \cite{bort} where the collision term is based on the particle 
vibration coupling model. Also in that model the damping rate strongly 
differs between low and high multipolarities of the collective motion.
However in ref.\cite{bort}, contrary to the spirit of the present work,
fast (diabatic) motion (the giant resonances) was considered, what makes 
a detailed comparison inadequate. In the future, in order to simulate a pure
mean field dynamics based on our model, we plan to calculate the evolution 
with the so called parallel ensemble technique and ensemble average at each 
instant of time the motion of the wall. We anticipate that this very much 
will suppress the particle wall damping mechanism so that the collective
oscillation will be mostly undamped, a feature which should bring the 
present study in closer contact with the one performed in ref.\cite{bauer}.

Our model certainly has implications beyond nuclear physics. it should be 
generic for all situations where a heavy particle (here the wall) is
moving in a Knudsen gas of light particles. For example the motion of a 
pendulum of mass M in a very rarefied gas of particles with mass
$m << M$ should show similar features as the ones studied within the 
present model. If the big particle can be approximated by a sphere suspended
on a spring and the gas is enclosed in a box, the whole system can be
considered as some sort of dynamic generalisation of a Sinai billiard.
The study of the damping of the heavy particle in such a situation would 
be particularly interesting.

\bigskip
\bigskip
\bigskip

{\bf {Acknowledgements}}

One of us (A.R.) would like to thank Vito Latora, Stefano Ruffo and Allan 
Lichtenberg for fruitful discussions.

\appendix

\section{}

Here we give some details about the two-dimensional wall formula.
Starting from the original formulation
given by Swiatecky in ref.\cite{wf}, after rescaling in two dimensions, 
we get at the following result 

\begin{equation}
\frac{dE}{dt} = \frac{4}{\pi} \rho \bar v \oint \dot q^2 dl + ....
(terms~ of~ higher~ order~ in~ \dot q^3)
\end{equation} 

$\rho$ and $\bar v$ are respectively the density and the average speed 
of the gas, $\dot q$ is the speed of the wall and $dl$ is the line 
element. By integration of eq.(A1), and neglecting corrections of order
$\alpha^2$ or higher, we easily get

\begin{equation}
\frac{dE}{dt} = 4 \rho \bar v  R_o^3 (\dot \alpha(t))^2
\end{equation} 

\noindent
where $\dot \alpha(t)$ can be calculated either without or with coupling.
In the former case $\alpha(t) = \alpha_o cos(\Omega t)$ and Eq.(A2) will read

\begin{equation}
\frac{1}{E_o}\frac{dE}{dt} = \frac{2 R_o}{\bar v} 
\Omega^2 {\alpha_o}^2 sin^2 (\Omega t)
\end{equation} 

\noindent
$E_o$ and $\alpha_o$ are the initial energy of the gas and the initial 
elongation. After more algebra we get the excitation energy of 
the gas $\Delta E / E_o$ 

\begin{equation}
\frac{\Delta E}{E_o} = \frac{R_o}{\bar v}~ 
\Omega~ {\alpha_o}^2~ [\Omega t - \frac{sin (2\Omega t)}{2}]
~~~~~~~~no~ coupling
\end{equation} 

In the case with coupling $\dot \alpha(t)$ comes directly from our numerical 
simulation. In fact, since the wall energy is

\begin{equation}
E_w = \frac{1}{2} M \dot \alpha^2 + \frac{1}{2} M \Omega^2 \alpha^2
\end{equation}

we easily get $\dot\alpha$ and substitute in Eq.(A2) obtaining

\begin{equation}
\frac{1}{E_o}\frac{dE}{dt} = \frac{R_o}{ \bar v} 
[\frac{2}{M} E_w - \Omega^2 \alpha^2]~~~~~~~with~ coupling
\end{equation} 
 
By integration on time we get again the excitation energy of the gas
$\Delta E / E_o$. The last equation is the time dependent
wall formula (TDWF).

In ref.\cite{wf2}, a generalized wall formula has been studied,
which takes into account for the increased average speed of the gas for 
large times. The new dissipation formula contains an additional term 
quadratic in time, besides the linear term (eq.A4). We have performed
similar calculations in two dimensions, both without and with 
coupling. The new formula for the excitation energy of the gas
reads 

\begin{equation}
\frac {\Delta E}{E_o} = 2 I(t) + \frac{3}{\pi} I^2(t)
\end{equation}

I(t) being respectively

\begin{equation}
I(t) = \frac {\alpha^2 \Omega R_o}{2 v} [\Omega t - \frac{sin(2 \Omega t)}
{2}]~~~~~~~~no~ coupling
\end{equation}

\begin{equation}
I(t) = \frac{R_o}{ \bar v} \int_0^t dt 
[\frac{2}{M} E_w - \Omega^2 \alpha^2]~~~~~~~with~ coupling
\end{equation}

\end{document}